\documentclass[aps,pre,twocolumn,groupedaddress,showpacs,floatfix]{revtex4}
\usepackage{graphicx}
\usepackage{dcolumn}
\usepackage{bm}
\usepackage{amssymb}
\usepackage{amsmath}
\usepackage{epsfig}

\def\LM#1#2{\left|\begin{array}{l}{#1}\\[1ex]{#2}\end{array}\right.}

\begin{document}
\title{The subdiffusive target problem: Survival probability}
\author{S. B. Yuste$^{1}$ and Katja Lindenberg$^{2}$}
\affiliation{$^{(1)} $Departamento de F\'{\i}sica, Universidad de
Extremadura, E-06071 Badajoz, Spain\\
$^{(2)}$ Department of Chemistry and Biochemistry 0340, and Institute
for Nonlinear Science,
University of California San Diego, 9500 Gilman Drive, La Jolla, CA
92093-0340, USA}

\date{\today}

\begin{abstract}

The asymptotic survival probability of a spherical target in the
presence of a single subdiffusive trap or surrounded by a sea of
subdiffusive traps in a continuous Euclidean medium is calculated.
In one and two dimensions the survival probability of the target in
the presence of a single trap decays to zero as a power law and as a power law with logarithmic
correction, respectively.
The target is thus reached with certainty, but it takes the trap an
infinite time on average to do so.
In dimensions higher than two a single trap may never reach the target and so the survival probability is finite
and, in fact, does not depend on whether the traps move diffusively or
subdiffusively.  When the target is surrounded by a sea of traps, on the
other hand, its survival probability decays as a stretched exponential
in all dimensions (with a logarithmic correction in the exponent for
$d=2$).  A trap will therefore reach the target with certainty, and will
do so in a finite time.  These results may be directly related to enzyme
binding kinetics on DNA in the crowded cellular environment.

\end{abstract}

\maketitle

\section{Introduction}
A number of recent experiments have pointed to the occurrence of
subdiffusive motion in biophysical
environments~\cite{kues,caspi,errington,weiss,GoldingCox06,palmer,wong,HalfordMarko04,xie},
that is, motion where the mean square displacement of
a ``walker" grows sublinearly in time, e.g.,
\begin{equation}
\langle r^2(t)\rangle \sim t^\gamma, \quad 0<\gamma<1.
\end{equation}
In particular, recent experiments on transport of large molecules in
living cells indicate that the crowded cellular environment leads
to such motion~\cite{kues,caspi,errington,weiss,GoldingCox06}.
The sublinear growth of the mean square displacement
comes about because the presence of a large number
of macromolecules leads to a medium that
may present impediments such as barriers, traps, or otherwise interrupted
pathways to the diffusive motions observed in
typically much sparser environments designed in vitro.  A number of
experimental and theoretical papers have
begun to address the problem of how to characterize the motion of
proteins in living cells and, having characterized it as subdiffusive,
of how to estimate the effects of this slowed-down transport on the
binding and unbinding of enzyme proteins on DNA target
sites~\cite{GoldingCox06,HalfordMarko04,Seki,LomholtZaidMetzler07}.
The results are not obvious and involve the balance
between reaching a target site more slowly (which might tend to slow
down reaction rates) while at the same time retaining the enzyme in the
vicinity of a target site for a longer time (which might tend to speed
it up).  Furthermore, the breaking of ergodicity indicates that one
might need to be careful about dealing with time histories of single
events vs ensemble averages over many events~\cite{LomholtZaidMetzler07}.
These issues arise not only in the biophysical context of cells
but in other situations involving crowded environments such as
porous media, but it is the cellular context that has awakened interest
in the problem most recently.  Various theories appear not to be entirely
consistent with one another in their predictions.
For example, Golding and Cox~\cite{GoldingCox06} predict singular
features in the binding/unbinding kinetics when the exponent $\gamma$
crosses the value $2/3$, while others~\cite{Seki,LomholtZaidMetzler07} do not
seem to predict such features.

The theoretical approaches to the problem are fairly complex and range
from continuous time random walk formalisms to scaling arguments.
The literature that focuses on the biophysical context includes a
number of realistic features such as the probability of a protein
unbinding from the target site on the DNA before a reaction takes place
and having to return to that site. These features add complications
surrounding the appropriate boundary conditions. The relation between
different boundary conditions used in this
problem~\cite{Seki,LomholtZaidMetzler07} are not entirely clear and
seem not to have been discussed in the literature.
Furthermore, it may be the case that the final results of interest
obey scaling laws that relate the more realistic problem to the simpler
one in which binding definitely leads to reaction~\cite{Seki}.
In any case, it seems to us that even the ``stripped-down" version of
the problem wherein a reaction occurs with certainty when the target
is reached would benefit from a systematic and clear presentation, and
this is our goal in this paper.

Given the scenario of a target surrounded by a sea of subdiffusive
particles the first of which to reach the target defines the binding
time of interest, what exactly is it that one wishes to calculate?  In
some instances, the quantity calculated is the probability that a
subdiffusive molecule initially at a distance $r$ from the target ever
finds it~\cite{GoldingCox06,Seki,Barkai01,Sung02}.
In other instances, it is the distribution
of the time-averaged probability of the subdiffusive molecule to be in
the bound state (vs the unbound state) in a single
trajectory~\cite{LomholtZaidMetzler07}.
A standard classic measure is the
distribution of first reaction times~\cite{Seki,Barkai01,Balakrishnan}.
We focus on a traditional quantity from which many
of these can in principle be deduced under appropriate
physical conditions, namely, the survival probability $Q(t)$
of the target as a function of
time~\cite{Sung02,BluZumKlaPRB84,BluKlaZuOptical}.
A specific scenario for this
calculation arises when a target (perhaps DNA) site
is surrounded by many randomly located particles (perhaps site-specific
DNA-binding proteins).  From the outset, we
focus on the problem in a spatial continuum and thus rely on the fractional
diffusion equation.  We note that essentially all the dynamical theories
on this topic in the literature either start from or arrive at a
continuum formulation, even those that begin with a continuous time
random walk, and we furthermore note that there seems to be little
argument about the fact that at least asymptotically and on a spatial
mesoscale such a formulation is appropriate.  The differences in the
literature lie not in the use of a fractional diffusion equation but
rather in the boundary conditions~\cite{Seki,LomholtZaidMetzler07,Sung02}.
In this context we note that while a
portion of the literature deals with partially absorbing targets, that
is, targets with a finite probability of letting the attached particle go
(to eventually return and to again be trapped or not), or equivalently,
with a finite rather than an infinite reaction rate when a particle
meets the target, we deal with the fully absorbing target. The
connection between the two types of targets may in many cases be
straightforward~\cite{Seki}.

The paper is organized as follows.
In Sec.~\ref{formalism} we present the formalism for the calculation of
the survival probability.  The results for all dimensions are presented
in Sec.~\ref{results}.  Finally, we conclude with a brief summary in
Sec.~\ref{conclusions}.

\section{Survival Probability: The Formalism}
\label{formalism}

We start by defining $Q_1(r,t;R)$ as the probability that a random
walker (``the particle") that started at location $\mathbf{r}$ at time
$t=0$ has survived until time $t$ in the presence of an absorbing sphere
of radius $R$ centered at the origin.  Here $r=|\mathbf{r}|$ in
recognition of the orientational symmetry of the problem.  The
probability $Q_1$ is the main quantity from which other results are
obtained.  If instead of one we have $N$ independent (sub)diffusive
particles, their combined survival probability $Q(t;R)$
is simply the probability that none of the $N$ particles has
entered the absorbing sphere.  If we assume that the particles are
randomly distributed in a volume $V$, then
\begin{equation}
Q(t;R)=\left[\frac{1}{V} \int_{r>R} d\mathbf{r}\,
Q_1(r,t; R)\right]^N,
\end{equation}
and if we further take the limits $N\to\infty$ and $V\to\infty$ holding
the density $\rho=N/V$ fixed, then this becomes
\begin{equation}
Q(t;R)=\exp\left\{-\rho \int_{r>R} d\mathbf{r} \,
[1-Q_1(r,t;R)]\right\}.
\label{seelater}
\end{equation}

The calculation of $Q_1(r,t;R)$ for subdiffusive particles can be
directly adapted from the corresponding calculation of this quantity for
diffusive
particles~\cite{SanoTachiyaJCP71,BarzykinTachiyaJCP99,BlytheBrayPRE03}.
We introduce the probability density $w(\mathbf{r}',t|\mathbf{r};R)$ that
the particle is at location $\mathbf{r}'$ at time $t$ if it started at
position $\mathbf{r}$ at $t=0$. As before, an absorbing sphere of radius
$R$ is centered at the origin $\mathbf{r}=0$. The survival probability
$Q_1(r,t;R)$ is related to this probability density by
\begin{equation}
Q_1(r,t;R)= \int d\mathbf{r}'\, w(\mathbf{r}',t|\mathbf{r};R).
\end{equation}
The probability density obeys the fractional diffusion equation together
with initial and boundary conditions,
\begin{align}
\frac{\partial}{\partial t} w(\mathbf{r}',t|\mathbf{r};R)&
=~_{0}D_{t}^{1-\gamma } \left[D \nabla_{\mathbf{r}'}^2w\right],
\label{ecdifuw}\\
w(\mathbf{r}',0|\mathbf{r};R)&=\delta(\mathbf{r}'-\mathbf{r}) ,\\
w(R,t|\mathbf{r};R)&=0 , \label{ccFrac} \\
\lim_{r'\to\infty}w(\mathbf{r}',t|\mathbf{r};R)&=0.
\label{bdry}
\end{align}
Here $\nabla_{\mathbf{r}'}$ is the Laplacian operator with respect to
the position $\mathbf{r}'$, $~_{0}\,D_{t}^{1-\gamma } $ is the
Riemann-Liouville operator
\begin{equation}
~_{0}\,D_{t}^{1-\gamma } f(\mathbf{x},t)=\frac{1}{\Gamma(\gamma)}
\frac{\partial}{\partial t} \int_0^t d\tau \,
\frac{f(\mathbf{x},\tau)}{(t-\tau)^{1-\gamma}},
\end{equation}
and $D$ is a generalized diffusion coefficient.

Laplace transforming Eqs.~(\ref{ecdifuw}), (\ref{ccFrac}), and
(\ref{bdry}) with respect to time according to
\begin{equation}
\widetilde{g}(u) = \int_0^\infty dt\, e^{-ut} g(t)
\end{equation}
yields for $\widetilde{w}(\mathbf{r}',u|\mathbf{r};R)$
\begin{align}
u \widetilde w -\delta(\mathbf{r}'-\mathbf{r}) &=u^{1-\gamma } D
\nabla_{\mathbf{r}'}^2
\widetilde w ,\label{first}\\
\widetilde w(R,u|\mathbf{r};R)&=0,\\
\lim_{r'\to\infty}\widetilde w(\mathbf{r}',u|\mathbf{r};R)&=0.
\end{align}
Note that $\widetilde{w}(\mathbf{r}',u|\mathbf{r};R)$ is the
Green function of Eq.~(\ref{first}). The reciprocity of the Green
function with respect to the arguments $\mathbf{r}'$ and $\mathbf{r}$
then implies that it also satisfies the adjoint equation
\begin{align}
u \widetilde w -\delta(\mathbf{r}'-\mathbf{r}) &=u^{1-\gamma } D
\nabla_{\mathbf{r}}^2
\widetilde w \label{wduN}
\end{align}
along with the boundary conditions~\cite{SanoTachiyaJCP71}
\begin{align}
\widetilde w(\mathbf{r}',u|R;R)&=0 , \label{wR0}\\
\lim_{r\to\infty}\widetilde w(\mathbf{r}',u|\mathbf{r};R)&=0.
\label{bdry2}
\end{align}
Integration over $\mathbf{r}'$ then gives for the Laplace transform
$\widetilde{Q}_1(r,u; R)$ of the quantity of interest
$Q_1(r,t;R)$~\cite{SanoTachiyaJCP71},
\begin{align}
u \widetilde {Q}_1 -1 &=u^{1-\gamma } D \nabla^2_{\mathbf{r}} \widetilde{Q}_1,
\label{Wecuu}\\
\widetilde{Q}_1(R,u;R)&=0 ,\label{tWR}\\
\lim_{r\to\infty}\widetilde Q_1(r,u;R)&=\frac{1}{u}  \label{tWinfi}.
\end{align}
Finally, the inverse Laplace transform yields the evolution equation and
boundary and initial conditions for the survival probability of the
particle that started its walk at $\mathbf{r}$,
\begin{align}
\frac{\partial }{\partial t}Q_1(r,t;R)&=~_{0}D_{t}^{1-\gamma }
\left[D \nabla_{\mathbf{r}}^2 Q_1\right],
\label{WecuT}\\
Q_1(R,t;R)&=0 ,\\
Q_1(r,t=0;R)&=1 ,\label{ccWt0}\\
\lim_{r\to\infty} Q_1(r,t;R)&=1.
\end{align}
The last condition is a result of the certain survival at any finite
time of a particle initially located at $r\to\infty$.
Due to the spherical symmetry of the problem, the $d$-dimensional
Laplacian operator is
\begin{equation}
\label{laplacianoperator}
\nabla_{\mathbf{r}}^2=\frac{\partial^2 }{\partial r^2}+\frac{d-1}{r} \frac{\partial }{\partial r}
\end{equation}
and the solution in Laplace space, valid for all $d$ (even for
non-integer $d$) is
\begin{equation}
\label{}
u \widetilde{Q}_1(r,u;R)=
   1-\left(\frac{r}{R}\right)^{1-d/2}
   \frac{K_{d/2-1}\left(\sqrt{r^2
u^\gamma/D}\right)}{K_{d/2-1}\left(\sqrt{R^2 u^\gamma/D}\right)},
\end{equation}
which will be analyzed in more detail subsequently.  Here $d$ is the
dimensionality and the $K$'s are modified spherical Bessel functions of
the third kind~\cite{Abramowitz}.
Although under some circumstances it may be useful and even illuminating
to consider non-integer values of $d$ (for example,
see~\cite{BlytheBrayPRE03}), our results are only physically relevant
for dimensions for which the the Laplacian operator~\eqref{laplacianoperator}
is meaningful, that is, for integer dimensions. As an aside, we note that this
subdiffusive result is related to the single particle survival probability
for a normal diffusive particle by the relation
\begin{equation}
\label{StTimeExpanded}
S(t|\gamma)=\int_0^\infty d\tau S(\tau|\gamma=1) T_\gamma(\tau,t)
\end{equation}
or, in Laplace space,
\begin{equation}\label{Suga}
\widetilde{S}(u|\gamma)=u^{\gamma-1} \widetilde{S}(u^\gamma|\gamma=1),
\end{equation}
where $S(t|\gamma)$ is the survival probability associated with a
physical situation involving a subdiffusive particle and
$S(t|\gamma=1)$ is the survival probability in the same physical
situation but involving a normally diffusive particle, and
$T_\gamma(\tau,t)$ is the ``time-expanding transformation'' \cite{SokoTET}
\begin{equation}
\label{}
\widetilde{T}_\gamma(\tau,u)=u^{\gamma-1} \exp\left(-\tau u^\gamma\right).
\end{equation}
In our case, Eq.~\eqref{Suga} means that $\widetilde Q_1(r,u;R|\gamma)=u^{\gamma-1}
\widetilde Q_1(r,u^\gamma;R|\gamma=1)$. [See~\cite{SokoTET,LaplaceScaling} for
more details on the integral transformation~\eqref{StTimeExpanded}
and the scaling relation~\eqref{Suga}.]

Before analyzing these results and using them to obtain explicit
survival probabilities, we note that one can provide an alternative
expression to Eq.~(\ref{seelater}) for the survival probability
$Q(t;R)$ in terms of $Q_1$.  For this purpose, we
define~\cite{BlytheBrayPRE03}
\begin{equation}\label{}
f(t;R)= \int_{r>R} d\mathbf{r}\,[1-Q_1(r,t;R)].
\end{equation}
Note that $f(0;R)=0$ because $Q_1(r,0;R)=1$.  Taking a time derivative of
this function,
\begin{equation}
\label{}
\frac{d}{dt}f(t;R)= -\int_{r>R} d\mathbf{r}\,
\frac{d}{dt}Q_1(r,t;R),
\end{equation}
and using Eq.~\eqref{WecuT}, we find that
\begin{equation}
\begin{aligned}
\label{}
\frac{d}{dt}f(t;R)&= -  \int_{r>R} d\mathbf{r} ~_{0}D_{t}^{1-\gamma }
\left[D \nabla_{\mathbf{r}}^2  Q_1(r,t;R)\right],\\
&=-D~_{0}D_{t}^{1-\gamma } \int_A d\mathbf{A} \cdot
\mathbf{\nabla}_{\mathbf{r}} Q_1(r,t;R),\\
&=D S_d R^{d-1} ~_{0}D_{t}^{1-\gamma } \left. \frac{\partial
}{\partial r}Q_1(r,t;R)\right|_{r=R},
\end{aligned}
\end{equation}
where $S_d=2\pi^{d/2}/\Gamma(d/2)$ is the surface of a sphere of
unit radius, and $d\mathbf{A}$ is a surface element in the direction
perpendicular to the surface of the sphere.
Since $f(0;R)=0$, we can write
\begin{equation}\label{}
f(t;R)=\int_0^t d\tau\,\frac{d}{d\tau}f(\tau;R)\\
=D S_d R^{d-1} F(t;R)
\end{equation}
with
\begin{equation}\label{Ftori1}
\begin{aligned}
F(t;R)&=\int_0^t
d\tau ~_{0}D_{\tau}^{1-\gamma }
\left.\frac{\partial}{\partial r} Q_1(r,\tau;R)\right|_{r=R},
\end{aligned}
\end{equation}
and consequently
\begin{equation}
\label{QTfrac2}
Q(t;R)
=\exp\left\{ -\rho D S_d R^{d-1} F(t;R)   \right\}.
\end{equation}
We will use this route in our calculations.  Our task is thus to
calculate $Q_1(r,t;R)$ and from it $F(t;R)$, to finally
arrive at the survival probability $Q(t;R)$.

\section{Survival Probability: Results}
\label{results}
In this section we present asymptotic results for the survival
probabilities in all dimensions.

\subsection{$d<2$}

For $d<2$ we find that as $u\to 0$
\begin{equation}\label{}
\widetilde{Q}_1(r,u;R)\sim\frac{\Gamma(d/2)(r^{2-d}-R^{2-d})}{2^{2-d} \Gamma(2-d/2)}
\frac{D^{d/2-1}}{u^{1-\gamma+\gamma d/2}},
\end{equation}
and using a Tauberian theorem this leads to
\begin{equation}\label{approximate}
\begin{aligned}
Q_1(r,t;R)&\sim \frac{\Gamma(d/2)}{2^{2-d} \Gamma(2-d/2) }
(r^{2-d}-R^{2-d}) \\
&\times \frac{D^{d/2-1}}{ \Gamma(1-\gamma+\gamma d/2)} t^{-\gamma+\gamma
d/2},  \quad t\to \infty.
\end{aligned}
\end{equation}
But
\begin{equation}\label{}
\begin{aligned}
&~_{0}D_{t}^{1-\gamma }\left. \frac{\partial}{\partial r}
Q_1(r,t;R)\right|_{r=R}\\
&~~~~\sim
\frac{\Gamma(d/2) (2-d)R^{1-d} D^{d/2-1}t^{-1+\gamma d/2}}{2^{2-d} \Gamma(2-d/2) \Gamma(\gamma d/2)}
,
\end{aligned}
\end{equation}
so that
\begin{equation}
\begin{aligned}\label{}
F(t;R)&=\int_0^t dt ~_{0}D_{t}^{1-\gamma } \left. \frac{\partial}{\partial
r} Q_1(\tau|r,R)\right|_{r=R}\\
&\sim \frac{2^{d-1} \Gamma(d/2) R^{1-d} D^{d/2-1} t^{\gamma d/2}. }{ \Gamma(1-d/2) \Gamma(1+\gamma d /2)}
\end{aligned}
\end{equation}
It then follows that
\begin{equation}
Q(t;R)
\sim\exp\left\{ -\rho \, \frac{\left(4 \pi D t^{\gamma}\right)^{d/2}}
{\Gamma(1-d/2)
\Gamma(1+\gamma d/2)} \right\}, \quad t\to\infty.
\label{onedim}
\end{equation}
Note that this is independent of the radius $R$ of the absorbing sphere.
Note also that whereas the survival probability
of the target in the presence of a single (sub)diffusive particle decays
as a power law with time, the decay law becomes a stretched exponential
when there are many particles at initially random locations.

With $d=1$ this coincides with the result we previously reported
in~\cite{YusteKatjaPRE05},
\begin{equation}\label{}
Q(t;R)\sim\exp\left\{ -\rho \frac{ \sqrt{4 D t^{\gamma}}
}{\Gamma(1+\gamma/2)}  \right\},
\quad t\to\infty.
\end{equation}
When in addition $\gamma=1$,
\begin{equation}\label{}
Q(t;R)\sim\exp\left\{ - \frac{ 4\rho \sqrt{Dt}}{\sqrt{\pi}} \right\},
\end{equation}
which coincides with the result reported in~\cite{BlytheBrayPRE03}.

Actually, for $d=1$ the survival probability of the target in the
presence of a single (sub)diffusive particle can be given exactly [and
the result~(\ref{approximate}) is then the asymptotic expansion of this
exact result] as follows:
\begin{equation}\label{Q1u}
\widetilde{Q}_1(r,u;R)= \frac{1}{u} -  \frac{\exp\left[\left(R-r \right)
\sqrt{u^\gamma/D}\right] }{u},
\end{equation}
so that
\begin{equation}\label{exact1d}
Q_1(r,t;R)= 1 -  H^{10}_{11}\left[\frac{r-R}{\sqrt{Dt^\gamma}}
    \LM{(1,\gamma/2)}{(0,1)}   \right].
\end{equation}
Here $H$ is a Fox function.
The probability density function $-dQ_1/dt$ of first passage times
corresponding to Eq.~\eqref{Q1u} agrees with the one previously obtained by
Balakrishnan~\cite{Balakrishnan} (see also Ref.~\cite{Barkai01}, where
$-dQ_1/dt$ is given in terms of a one-sided L\'evy stable density).
 When $\gamma=1$ the Fox function reduces to
the complementary error function and Eq.~(\ref{exact1d}) then reduces to
the classic diffusive result~\cite{BarzykinTachiyaJCP99}.

\subsection{$d=2$}

In two dimensions the solution of the fractional diffusion equation with
the appropriate initial and boundary conditions leads to
\begin{equation}
\begin{aligned}\label{}
\widetilde{Q}_1(r,u;R)&\sim\frac{1}{u} \frac{\ln (R/r)}
  {\gamma_{\text{E}}+ \ln (R \sqrt{u^{\gamma}/(4D)})}\\
&\sim\frac{1}{u} \frac{\ln (R/r)}
  {\ln (R \sqrt{u^{\gamma}/(4D)})},\quad u\to 0,
  \end{aligned}
\end{equation}
where $\gamma_{\text{E}}\simeq 0.577216$ is the Euler-Mascheroni constant.
Applying the Tauberian theorem then leads to the asymptotic result
\begin{equation}
\label{2d}
Q_1(r,t;R)\sim \frac{2  \ln(r/R)}{\ln(4Dt^\gamma/R^2)} \equiv
\frac{2\ln(r/R)}{\gamma\ln(at)}.
\end{equation}
To calculate $F(t;R)$ we need to evaluate the fractional derivative
$G(t)\equiv ~_{0}D_{t}^{1-\gamma } g(t)$, with $g(t)\equiv 1/\ln(at)$.
We know that
\begin{equation}\label{}
\widetilde G(u) =  u^{1-\gamma}\widetilde g(u)
-\lim_{t\to 0} \frac{1}{\Gamma(\gamma)} \int_0^t ds \frac{g(s)}{(t-s)^{1-\gamma}}.
\end{equation}
However, the limit term on the right can easily be shown to vanish.
Since $\widetilde{g}(u) \sim [u\ln(a/u)]^{-1}$ as $u\to 0$, it follows that
$\widetilde{G}(u) \sim [u^\gamma\ln(a/u)]^{-1}$ and consequently
$G(t)\sim t^{\gamma-1}/[\Gamma(\gamma)\ln(at)]$ as $t\to \infty$. We
thus have that
\begin{equation}
F(t;R)=\frac{2}{R\gamma}\int_0^t d\tau G(\tau)\sim \frac{2 t^\gamma }{R
\gamma^2 \Gamma(\gamma) \ln(at) }
\end{equation}
and consequently
\begin{equation}\label{QTfracA}
Q(t;R)=\exp\left\{ -\rho   \frac{4\pi D t^\gamma}{\Gamma(1+\gamma)
\ln(4Dt^\gamma/R^2) }   \right\}, \quad t\to\infty.
\end{equation}
While the decay of the survival probability of a single particle is thus
an inverse logarithm, that of the ensemble of particles is a stretched
exponential with a logarithmic correction.  Also, when $d=2$ [and also
when $d>2$(see below)] the survival probabilities do depend on the
radius $R$ of the absorbing sphere.

\subsection{$d>2$}

For $d>2$ we have that
\begin{equation}\label{}
\widetilde{Q}_1(r,u;R)\sim\frac{1}{u} \left[ 1-\left(\frac{R}{r}\right)^{d-2}
\right],\quad u\to 0,
\end{equation}
so that
\begin{equation}\label{3d}
Q_1(r,t;R)\sim \left[ 1-\left(\frac{R}{r}\right)^{d-2} \right] ,\quad t\to
\infty.
\end{equation}
From this we obtain
\begin{align}\label{}
F(t)&=\int_0^t dt ~_{0}D_{t}^{1-\gamma } \left. \frac{\partial}{\partial
r} Q_1(r,\tau;R)\right|_{r=R}\\
&\sim \frac{d-2}{R} \frac{t^{\gamma}}{\Gamma(1+\gamma)} ,\quad t\to
\infty
\end{align}
from which it follows that
\begin{equation}\label{3dall}
Q(t;R)=\exp\left\{ -\rho \,\frac{S_d R^{d-2} (d-2) }{\Gamma(1+\gamma)}D
t^{\gamma} \right\}, \quad t\to\infty
\end{equation}
in agreement with results reported in \cite{Sung02,BluKlaZuOptical}. The survival probability of a single particle approaches a constant at
long times, whereas the ensemble survival probability decays as a
stretched exponential.

As in the one-dimensional case, for $d=3$ it is again possible to
provide an exact survival probability of the target in the presence of a
single (sub)diffusive particle as follows:
\begin{equation}\label{Q1ud3}
\widetilde{Q}_1(r,u;R)= \frac{1}{u} -\frac{R}{r} \frac{ e^{ \left(R-r \right)
\sqrt{u^\gamma/D} } }{u},
\end{equation}
so that
\begin{equation}\label{Q1td3}
Q_1(r,t;R)= 1 -\frac{R}{r} H^{10}_{11}\left[\frac{r-R}{\sqrt{Dt^\gamma}}
\LM{(1,\gamma/2)}{(0,1)}   \right].
\end{equation}
This Fox $H$ function reduces to a complementary error function
when $\gamma=1$, and in this limit we recover the classic normal
diffusion result, see e.g.~\cite{BarzykinTachiyaJCP99}.
It is interesting to note that for $d=3$ this result is the same as Eq.
(5.5) in~\cite{Seki} if we take the completely absorbing limit
$k_\gamma\to\infty$ in that formula.
Note that the probability density function $-dQ_1/dt$ of first passage times
corresponding to Eqs.~\eqref{Q1ud3} and \eqref{Q1td3} agrees with the one
previously obtained by Barkai~\cite{Barkai01}.

\section{Conclusions}
\label{conclusions}

We have calculated the asymptotic survival probability of an absorbing
target of radius $R$ at the origin in the presence of
one, or of many, (sub)diffusive particles.
This calculation, which is based on
the fractional diffusion equation, has been carried out for all dimensions,
with results that in some cases and
limits agree with
known results. Equations~ (\ref{approximate}) and
(\ref{onedim}) give these survival probabilities for $d=1$, Eqs.~(\ref{2d})
and (\ref{QTfracA}) for $d=2$, and Eqs.~(\ref{3d}) and (\ref{3dall}) for
$d\ge 3$.  Thus, in one dimension the survival probabilities are
respectively of power law and stretched exponential form, in two
dimensions the decay is slower, respectively logarithmic and stretched
exponential with a logarithmic correction.  The result in dimensions
three~\cite{Seki} and higher are interesting: the survival probability
of a single particle goes asymptotically to a constant (i.e., the particle
may survive forever with a finite probability), and this probability is
independent of the subdiffusive exponent, and thus the same as for a
normally diffusive particle.
The survival probability of the target surrounded by a sea of
subdiffusive traps does decay, again as a stretched exponential that
does depend on the subdiffusive exponent.
Note that in all cases, while the mean survival time
of the target in the presence of a single particle (i.e., the first
moment of $Q_1$) is infinite, that of a target in a large (infinite)
volume containing a finite density of particles (i.e., the first moment
of $Q$) is finite.  This may be an interesting observation
beyond the biophysical examples mentioned in the introduction, for example
in the search by an enzyme of a DNA target site involving a combination of
scanning and relocation events in which the relocation times have a
power law distribution with diverging moments.  The mean survival time
when a single enzyme seeks the target in this three-dimensional search
diverges (the survival probability of the target is a power law
without moments), but one of many enzymes will reach the target with
certainty (the survival probability of the target is a stretched
exponential).

The quantities that we have calculated lead immediately to others
frequently used in the literature as well as to additional insights.  For
example, the derivatives $-\partial Q_1/\partial t$ and
$-\partial Q/\partial t$ are the distributions of first passage times to
the absorbing target.  It is also noteworthy that the probability that a
subdiffusive molecule initially at a distance $r$ from the target ever
reaches it is the same for normally diffusive and subdiffusive
particles.  In particular, since $\lim_{t\to\infty}
Q_1(r,t;R)=Q_1(r,\infty;R)$ vanishes for
$d=1$ and $d=2$, a (sub)diffusive particle will eventually reach the
target with certainty. On the other hand, since for $d\ge 3$ this limit
is finite, $1-(R/r)^{d/2}$, a particle escapes the target forever
with probability $(R/r)^{d/2}$ whether it is diffusive or subdiffusive.
When the target is surrounded by a sea of particles, however,
since $\lim_{t\to\infty} Q(r,t;R)=Q(r,\infty;R)$ vanishes for all
dimensions, one of the particles, whether diffusive or subdiffusive,
eventually reaches the target with certainty.  The
approach to these asymptotic behaviors of course depends on
dimensionality and also on the subdiffusive exponent $\gamma$.

Our immediate future work on this topic will focus on the effects of a
partially absorbing target, that is, a target that does not necessarily
``die" upon its first encounter with a (sub)diffusive particle.  Some
results on this case have been reported in the literature, notably
in~\cite{Seki} and~\cite{LomholtZaidMetzler07}, but they do not use the
same boundary conditions and they also do not consider this situation in
all dimensions.  Clearly, there is still work to be done.

\section*{Acknowledgments}
The research of S.B.Y. has been supported by
the Ministerio de Educaci\'on y Ciencia (Spain) through grant
No. FIS2007-60977 (partially financed by FEDER funds). K.L. is supported in part by the
National Science Foundation under grant PHY-0354937.

\end{document}